\pdfoutput=1
\documentclass[10pt,conference]{IEEEtran}
\IEEEoverridecommandlockouts
\usepackage{cite}
\usepackage{amsmath,amssymb,amsfonts}
\usepackage{graphicx}
\usepackage{xcolor}
\def\BibTeX{{\rm B\kern-.05em{\sc i\kern-.025em b}\kern-.08em
    T\kern-.1667em\lower.7ex\hbox{E}\kern-.125emX}}

\def\HiLi{\leavevmode\rlap{\hbox to \hsize{\color{yellow!50}\leaders\hrule height .8\baselineskip depth .5ex\hfill}}}

\usepackage{pifont}
\usepackage{mathtools}
\usepackage[inline]{enumitem}
\usepackage{caption}
\usepackage{booktabs}
\usepackage{array}
\usepackage{multirow}
\usepackage{xspace}
\usepackage{letltxmacro}
\usepackage[linesnumbered,ruled,vlined,procnumbered]{algorithm2e}
\usepackage{etoolbox}

\makeatletter
\AtBeginEnvironment{procedure}{\let\c@algocf\c@procedure}
\makeatother

\newcommand{\subscript}[2]{$#1 _ #2$}

\DeclareMathOperator{\E}{\mathbb{E}}

\definecolor{purple}{rgb}{1, 0, 1}

\newcommand{\ie}{\emph{i.e.,}\xspace}
\newcommand{\eg}{\emph{e.g.,}\xspace}

\newcommand{\ea}{\emph{et al.}\xspace}

\newcommand{\srep}{\emph{SREP}\xspace} 
\newcommand{\srepsim}{\emph{SREPSim}\xspace}
\newcommand{\esrep}{\emph{E-SREP}\xspace}
\newcommand{\epsrep}{\emph{EP-SREP}\xspace}

\newcommand{\mempoolsync}{\emph{MempoolSync}}

\newcommand{\fref}[1]{Fig.~\ref{#1}}
\newcommand{\tref}[1]{Table~\ref{#1}}
\newcommand{\aref}[1]{Algorithm~\ref{#1}}
\newcommand{\procref}[1]{Procedure~\ref{#1}}
\newcommand{\sref}[1]{Section~\ref{#1}}
\newcommand{\lineref}[1]{line~\ref{#1}}

\LetLtxMacro{\originaleqref}{\eqref}
\renewcommand{\eqref}{Eq.~\originaleqref}

\newcounter{theoremcount}
\DeclareRobustCommand{\theorem}[1]{%
  \refstepcounter{theoremcount}%
  \noindent\textit{\textbf{Theorem \thetheoremcount\label{theorem:#1}: }}%
}
\DeclareRobustCommand{\theoremref}[1]{Theorem~\ref{theorem:#1}}

\DeclareRobustCommand{\proof}{\emph{Proof:}\xspace}
\DeclareRobustCommand{\qqed}{\hfill$\blacksquare$}

\newcounter{corollcount}
\DeclareRobustCommand{\coroll}[1]{%
  \refstepcounter{corollcount}%
  \noindent\textit{\textbf{Corollary \thecorollcount\label{coroll:#1}: }}%
}

\newcounter{lemmacount}
\DeclareRobustCommand{\lemma}[1]{%
  \refstepcounter{lemmacount}%
  \noindent\textit{\textbf{Lemma \thelemmacount\label{lemma:#1}: }}%
}
\DeclareRobustCommand{\lemmaref}[1]{Lemma~\ref{lemma:#1}}

\newcounter{definitioncount}
\DeclareRobustCommand{\definition}[1]{%
  \refstepcounter{definitioncount}%
  \noindent\textit{\textbf{Definition \thedefinitioncount\label{definition:#1}: }}%
}
\DeclareRobustCommand{\defref}[1]{Definition~\ref{definition:#1}}

\newif\ifnotes
\notestrue
\notesfalse

\newif\ifdiff
\difftrue
\difffalse

\IEEEoverridecommandlockouts

\begin{document}


\title{\srep{}: Out-Of-Band Sync of Transaction Pools for
  Large-Scale Blockchains}

\author{\IEEEauthorblockN{Novak Boškov, Sevval Simsek, Ari
    Trachtenberg, and David Starobinski}
  \IEEEauthorblockA{\textit{Department of Electrical and Computer
      Engineering}\\
    \textit{Boston University, Boston, Massachusetts, USA}\\
    \{boskov,sevvals,trachten,staro\}@bu.edu}
}


\IEEEoverridecommandlockouts

\IEEEpubid{\makebox[\columnwidth]{979-8-3503-1019-1/23/\$31.00~\copyright2023 IEEE \hfill} \hspace{\columnsep}\makebox[\columnwidth]{ }}

\maketitle

\IEEEpubidadjcol

\begin{abstract}
Synchronization of transaction pools (\emph{mempools})  has shown
potential for improving the performance and block propagation delay
of state-of-the-art blockchains.
Indeed, various heuristics have been proposed in the literature to this end, all of
which incorporate exchanges of unconfirmed transactions into their block
propagation protocol. In this work, we take a different approach, maintaining
transaction synchronization outside (and independently) of the block propagation
channel.  In the process, we formalize the synchronization problem within a graph
theoretic framework and
introduce a novel algorithm (\srep{} - \emph{Set Reconciliation-Enhanced Propagation})
with quantifiable guarantees. We analyze the algorithm's performance for various realistic network topologies,
and show that it converges on any connected graph in a number of steps that is bounded by the
diameter of the graph. We confirm our analytical findings through extensive simulations that include
comparison with \mempoolsync{}, a recent approach from the literature. Our simulations show that
\srep{} incurs reasonable overall bandwidth overhead and, unlike \mempoolsync{}, scales gracefully with the
size of the network.

\end{abstract}



\section{Introduction and Related Work}
Block propagation represents a fundamental aspect of many blockchain
networks in which blockchain nodes forward newly created blocks to
their neighbors. Historically, block propagation has been performed by
sending all the transactions belonging to the block alongside the
block's metadata. Often, a substantial number of the block's
transactions are present on the receiving end, resulting in
unnecessarily high \emph{bandwidth overhead}. To cope with such
overhead, more advanced block propagation protocols such as
\emph{CompactBlock}~\cite{compact_block}, \emph{Xtreme Thin
  Blocks}~\cite{xthin}, \emph{Graphene}~\cite{ozisik2019graphene}, and
\emph{Gauze}~\cite{ding2022} have been introduced. 

Yet, it has recently been demonstrated through \emph{in-situ}
measurements in live blockchains, including Bitcoin, that the
performance of these advanced block propagation protocols can
significantly degrade when transaction pools go out of
sync~\cite{anas_empir,anas_churn_tnsm,anas_churn_icbc,churn_misic}.
One approach to prevent such performance degradation is to have
neighboring nodes regularly synchronize their pools of unconfirmed
transactions. Toward this end, the recent work
in~\cite{anas_churn_tnsm} proposes a heuristic, called \mempoolsync{},
that is shown to reduce the average block propagation delay by~50\% in
the Bitcoin network. Yet, \mempoolsync{} does not provide any
quantifiable \emph{guarantees} on overall communication or delay
performance.

In this work, we study the problem of transaction pool synchronization
(\emph{sync}) from a fundamental, graph-theoretic perspective, which
allows us to analyze synchronization performance metrics in various
network topologies. Our main contributions are as follows:
\begin{itemize}
\item We introduce a novel transaction pool sync algorithm, called
  \srep{}, which functions in an assistive capacity
  outside of the existing block propagation protocols.
\item We analyze the performance of \srep{} in general network
  topologies, including a more specialized model that captures
  topological properties of actual blockchains (\eg{} the
  ``small-world'' property) as well as the statistics of transaction
  pools.
\item We develop a simulation approach based on realistic transaction
  pool data from measurement campaigns, and confirm our analytical
  findings through simulations.
\item We show that \srep{} has significantly lower bandwidth
  overhead than \mempoolsync{}.
\end{itemize}

The rest of this paper is organized as follows. In \sref{sec:rw}, we
overview the related work. In \sref{sec:srep_algo}, we introduce
\srep{}. In \sref{sec:analysis}, we analyze the properties of \srep{}
and validate our findings through simulations in \sref{sec:sim}. We
compare \srep{} with a transaction pool synchronization approach from
the literature in \sref{sec:mempoolsync}. Finally, we give a
conclusion and propose future work in \sref{sec:conclusion}.


\section{Background}\label{sec:rw}

To the best of our knowledge, \srep{} is a unique distributed
algorithm that explicitly tackles the problem of network-wide
synchronization of unconfirmed transactions --- \emph{transaction
  pools}~\cite{blockchain_consensus_survey}. To achieve its goals,
\srep{} relies on \emph{communication-efficient} solutions to the
\emph{set reconciliation} problem~\cite{minsky2002practical}, which is
defined as follows. Given two remote parties with their corresponding
data sets $S_A$ and $S_B$, each party needs to discover the elements
local to the other. Communication-efficient solutions to this problem
exchange only messages of size proportional to the number of
\emph{mutual differences} defined as
$(S_A \setminus S_B) \cup (S_B \setminus S_A)$ and often denoted as
$S_A \oplus S_B$.

In fact, there has been several communication-efficient set
reconciliation algorithms proposed in the literature including
Characteristic Polynomial Interpolation~\cite{minsky2003set} (CPI),
BCH codes~\cite{dodis2004fuzzy}, and Invertible Bloom Lookup Tables
(IBLT)~\cite{goodrich2011invertible,eppstein2011s,iblt_new}. For
instance, CPI incurs a communication cost \emph{equal} to the number
of mutual differences plus a small constant, which makes it nearly
\emph{optimal} in communication~\cite{minsky2003set}. On the other
hand, IBLT-based solutions typically offer better \emph{computational}
complexity at the cost of increasing their communication cost by a
constant factor. To further reduce this communication overhead, Lázaro
and Matuz~\cite{iblt_new} have recently proposed an IBLT-based
solution that brings the communication cost closer to that of CPI
while keeping the computational complexity low.



On the other hand, when it comes to our analytical model and
simulations, we make use of the findings from the blockchain
topology-discovering literature. In particular,
Wang~\ea{}~\cite{ethna} and Gao~\ea{}~\cite{mes_ether_topo}
independently verified that the Ethereum network exhibits
``small-world'' property. Recently,
Shahsavari~\ea{}~\cite{theory_model} used a random graph model to
simulate Bitcoin network and Ma~\ea{}~\cite{cblocksim} proposed a
topology generation based on Watts-Strogatz~\cite{Watts1998} random
graph model to capture the Bitcoin network in their \emph{CBlockSim}
simulator.

\section{SREP Algorithm}\label{sec:srep_algo}
We propose a novel distributed algorithm for network-wide transaction
pool synchronization called \srep{} (\emph{Set Reconciliation-Enhanced
  Propagation}). The core building block of \srep{} is a concept that
we denote as \emph{primal} sync --- a set reconciliation protocol with
communication complexity linear in the number of symmetric differences
(\eg{} CPI~\cite{minsky2003set}). Given the local transaction pool as
a set of globally unique identifiers~\cite{utxo}, \srep{} invokes one
primal sync per each neighbor in parallel.

\begin{figure*}
  \centering
  \includegraphics[width=\linewidth]{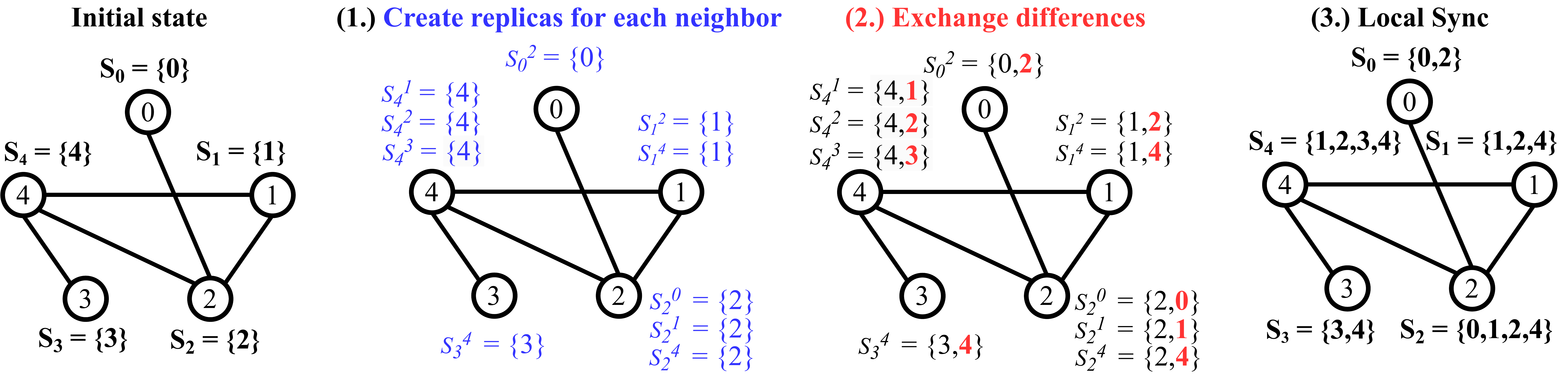}
  \caption{One iteration of \srep{} on a tractably small
    network.}
  \label{fig:parallel_srep}
\end{figure*}

One way to support many parallel invocations of primal syncs is to
create one transaction pool \emph{replica} per each neighbor. Then run
primal syncs in parallel using the corresponding replicas to avoid
write collisions. Upon the completion of all parallel tasks, we can
reuse the primal sync to incorporate new elements into the local
transaction pool. We describe \srep{} in \aref{algo:parallel_srep}
using $S_{n}$ to denote the transaction pool at node $n$, $d_{in}$ to
denote the differences between $S_i$ and $S_n$ that reside in $S_i$,
and \textbf{Sync} to denote a primal sync. As an illustration, in
\fref{fig:parallel_srep}, we depict one iteration of \srep{}'s main
loop (\lineref{line:srep_main_loop}), assuming that each node $n$
holds only one transaction whose hash is also $n$.


\begin{algorithm}
  \small
  \caption{\srep{} Algorithm.}\label{algo:parallel_srep}
  \SetKwBlock{Ateach}{At each node $n \in \{0, |V| - 1\}$}{end}
  \SetKwBlock{InParallel}{Do in parallel}{end}
  \SetKwBlock{Loop}{Loop}{EndLoop}
  \SetKw{kwSpawn}{spawn thread}
  \SetKw{kwSync}{Sync}
  \SetKw{kwForIn}{in}
  \KwIn{Network $G = (V, E)$ as adjacency list.}
  \Ateach{
    \Loop { \label{algo:parallel_srep:loop}\label{line:srep_main_loop}
      \For(\tcp*[h]{Neighbors of $n$}){$i$ \kwForIn $G[n]$} {
        $S_n^i \gets S_n$ \tcp*[l]{Replicate data set}
        \InParallel{
            \tcp{Network sync}
            $d_{in} \gets$ \kwSync( $S_n^i$, $S_i$ ) \;
            $S_n^i \gets S_n^i \cup d_{in}$ \;
        }
      }
      \For{$i$ \kwForIn $G[n]$} {
        \tcp{Local sync}
        $S_n^i \setminus S_n \gets$ \kwSync( $S_n$, $S_n^i$ ) \;
        $S_n \gets S_n \cup (S_n^i \setminus S_n)$ \;
      }
    }
  }
\end{algorithm}

\subsection*{Avoiding Full Replication}

\srep{} from \aref{algo:parallel_srep} has a significant memory
overhead caused by transaction pool replication for each
neighbor. However, certain primal syncs allow us to implement \srep{}
without replication, thus mitigating this memory overhead. In
particular, multiple set reconciliation algorithms mentioned in
\sref{sec:rw} use data set \textit{sketches} to perform
synchronization and modify the underlying data sets only at the end of
the protocol.

For instance, CPI reads from the set only once, at the beginning of
the protocol, and writes to it only once at the end of the protocol.
Suppose that we choose CPI as the primal sync in \srep{}. Then we can
construct the characteristic polynomial~\cite{minsky2002practical} of
$S_n$ as the very first step in each iteration (after
\lineref{algo:parallel_srep:loop} in
\aref{algo:parallel_srep}). Instead of using the neighbor replicas, we
can now use the same characteristic polynomial in all neighbor
threads.
As no thread will modify the polynomial, the procedure is thread-safe
and the threads can now write directly to the underlying set.
Although the write operation will need to acquire the corresponding
lock, since set union is commutative and associative, the order in
which the threads update the set does not matter. 
As we now avoid replication, the local synchronization step can be
safely eliminated altogether.

Note that this implementation improvement does not change the
functional properties of \srep{}. That is, each thread still operates
on its own version of the sketch and will update its sketch only at
the beginning of the subsequent iteration. Hence, a difference that
arrives in iteration $i$ via some neighbor thread will only get
acknowledged by other threads in iteration $i + 1$. For that reason,
we use the notion of ``replicas'' in the subsequent analysis.

\begin{table}
  \centering
  \begin{tabular}{>{\centering\arraybackslash}m{2cm} m{5cm}}
    \toprule[1pt]
    $G = (V, E)$ & Network of $|E|$ edges and $|V|$ nodes \\ \addlinespace[.3em]
    $S_n$ & Transaction pool at node $n \in \{0..|V| - 1\}$ \\ \addlinespace[.3em]
    $d_{ij} = S_i \setminus S_j$ & Differences between $i$ and $j$ that
                                   reside in $i$\\ \addlinespace[.3em]
    $\overline{deg}$ & Average node degree \\ \addlinespace[.3em]
    $t_n$ & Time node $n$ spends to synchronize with all its neighbors once \\ \addlinespace[.3em]
    $T_{x\%}$ & Time until $x$\% of $G$ is synchronized \\ \addlinespace[.3em]
    $\Sigma_{x\%}$ & Number of primal sync invocations \\ \addlinespace[.3em]
    $C_{x\%}$ & Overall communication cost \\
    \bottomrule[1pt]
  \end{tabular}
  \caption{Summary of notation.}
  \label{tab:not}
\end{table}

\section{\srep{} Performance Analysis}\label{sec:analysis}
Several aspects affect the performance of \srep{}, including the
network topology and the statistics of transaction pools. To aid our
analysis, we first define an explicit network model, and then analyze
\srep{} in a step-by-step fashion. In each stage of our analysis, we
describe a \srep{} variant with the corresponding set of
\emph{simplifying assumptions} and analyze its performance. By
successively relaxing these assumptions, we arrive at the final
version of \srep{}. \tref{tab:not} summarizes notation used throughout
this work.

\definition{percentage} We use $T_{x\%}$, $\Sigma_{x\%}$, and
$C_{x\%}$ to denote time, total number of primal sync invocations, and
total communication cost until $x\%$ of transaction pools in the
network are equal. When $x = 100$, we say that \emph{full network}
synchronization is achieved --- the ultimate goal of \srep{}.


\subsection{Network Model}\label{sec:net_model}
Watts-Strogatz~\cite{Watts1998} random graphs allow us to describe a
wide range of realistic blockchain network topologies reasonably
well~\cite{mes_ether_topo, ethna, under_the_hood, cblocksim}. A
typical set of parameters to Watts-Strogats model are the number of
nodes in the network $|V|$, average node degree $\overline{deg}$, and
rewire probability $p$~\cite{Watts1998}.

For instance, each Bitcoin node selects 8 random neighbors upon
joining the
network~\cite{bitcoin,txprobe,degwithunreachable},
which has been shown to yield an unstructured random
graph~\cite{theory_model}. We can capture this in the Watts-Strogatz
model by setting $\overline{deg} = 8$ and $p = 1$.  Ethereum's
neighbor selection mechanism, on the other hand, relies on a Kademlia
distributed hash table (DHT)~\cite{kademlia}, and yields a network
with more structure~\cite{mes_ether_topo}. Notwithstanding this,
multiple recent measurement results have independently confirmed that
the generated network exhibits the ``small world'' property and fits
the Watts-Strogatz model~\cite{mes_ether_topo,
  ethna,under_the_hood}. That is, the average shortest path between
any two nodes can be reasonably approximated by
$O\left( log_{\overline{deg}}|V| \right)$, and the diameter of the
network is small~\cite{chung2001diameter}.





Besides the graph topology, our network model also captures the states
of transaction pools across the network. In particular, we define the
\emph{pool assignment} $A$ as a collection of sets
$S_{0}..S_{|V| - 1}$ where set $S_{i}$ represents the transaction pool
at node $i$. We model the statistical properties of $A$ through the
following \emph{pool parameters}:
\begin{enumerate}
\item[$\mathcal{S}$:] \emph{sizes distribution}. A discrete random
  variable describing the sizes of transaction pools $S_{i}$ for
  $i \in \{0...|V| - 1\}$,
\item[$s$:] \emph{sizes vector}. A $|V|$-size vector where
  elements are drawn from $\mathcal{S}$,
\item[$\mathcal{P}$:] \emph{differences distribution}. A discrete
  random variable describing the sizes of mutual differences between
  the pairs of transaction pools (\ie{} $|S_{i} \oplus S_{j}|$),
\item[$M$:] \emph{mutual differences matrix}. A $|V| \times |V|$ upper
  triangular matrix of mutual differences. For the given topology
  $G = (V, E)$, the elements of the matrix are defined as:
  \[
    m_{ij} =
    \begin{dcases}
      |S_{i} \oplus S_{j}| & \text{when } (i, j) \in E \text{ and } i < j,\\
      0 & \text{otherwise}.
    \end{dcases}
  \]
  Non-zero elements are drawn from $\mathcal{P}$.
\item[$\mathcal{U}$:] \emph{universe}. A discrete random variable
  from which we draw transaction IDs. We choose $\mathcal{U}\{0, u\}$
  to be a uniform random variable for some $u \geq |V|$.
\end{enumerate}


\subsection{Elementary \srep{} (\esrep{})}
The starting point for our build up of \srep{} is called \emph{elementary} \srep{}
(\aref{algo:elem_srep}). We summarize its simplifying assumptions as
follows:
\begin{enumerate}[label=(\subscript{A}{{\arabic*}})]
\item All nodes have global view of the network.\label{a:global_view}
\item Initially, the transaction pools at each node contain only one
  element (transaction) that is unique across all network nodes (\eg{}
  index of the node). Strictly speaking, we set the pool parameters
  as: $\mathcal{S} = 1$, $\mathcal{P} = 2$ 
  , and
  $u \gg |V|$.\label{a:single_elem}
\item No new transactions arrive to the network after the
  initialization.\label{a:data_gen}
\item In one iteration of elementary \srep{}
  (\lineref{line:elem_srep_iter}), nodes take turns to perform their
  synchronization duties such that no two nodes invoke primal sync at
  the same time. For instance, nodes with smaller indices go first. An
  iteration ends when all nodes have invoked synchronization once for
  all their neighbors.\label{a:precedence}
\item Nodes synchronize with their neighbors sequentially. For
  instance, the neighbors with smaller indices get synchronized first
  (\lineref{line:elem_srep_sort}).\label{a:local_order}
\item All synchronizations are two-way (lines~\ref{line:sync_one_way}
  and~\ref{line:sync_other_way}), meaning that the differences are
  exchanged in both directions.\label{a:two_way}
\item All synchronizations take equally long.\label{a:equal_duration}
\end{enumerate}

\begin{algorithm}
  \small
  \caption{Elementary \srep{}.}\label{algo:elem_srep}
  \SetKwBlock{Ateach}{At each node $n \in \{0, |V| - 1\}$}{end}
  \SetKw{kwSync}{Sync}
  \SetKw{kwForIn}{in}
  \SetKw{kwNot}{not}
  \SetKw{kwSort}{sort}
  \KwIn{Network $G = (V, E)$ as adjacency list.}
  \While {network is \kwNot fully synchronized} { \label{line:elem_srep_iter}
    \For{$n \gets 0$ \KwTo $\{0..|V| - 1\}$} {
      $neighbors \gets$ \kwSort( $G[n]$ ) \; \label{line:elem_srep_sort}
      \For{$i$ \kwForIn $neighbors$} {
        $d_{in} \gets$ \kwSync( $S_n$, $S_i$ ) \;
        $d_{ni} \gets$ \kwSync( $S_i$, $S_n$ ) \;
        $S_{n} \gets S_{n} \cup d_{in}$ \; \label{line:sync_one_way}
        $S_{i} \gets S_{i} \cup d_{ni}$ \; \label{line:sync_other_way}
      }
    }
  }
\end{algorithm}

In the context of \esrep{}, the following special case is particularly
significant for the analysis.

\lemma{elemsrepcomplete} For \esrep{} over a complete graph
$G = (V, E)$, the communication cost to sync the entire network is
\[
  C_{100\%}(G) = |V| \cdot (|V| - 1).
\]

\subsection{Elementary Parallel \srep{} (\epsrep{})}\label{sec:p_srep}
The main aim of the \textit{elementary parallel} \srep{} is to relax
\ref{a:global_view}, \ref{a:precedence} and
\ref{a:local_order}. Instead of invoking synchronization in order,
\epsrep{} invokes synchronization for all neighbors at once (\ie{}
\aref{algo:parallel_srep}). In addition to that, we also relax
\ref{a:equal_duration}. The synchronization between nodes $u$ and $v$
now takes time \emph{equal} to the number of their mutual differences
(\ie $|d_{uv} \cup d_{vu}|$). As discussed earlier in \sref{sec:rw},
this is a reasonable assumption to make (\eg{} CPI has such a
property).


\theorem{comm_bounds} In \epsrep{} and for any connected network
$G = (V, E)$, we have the following bounds on the overall
communication cost until the network is fully synchronized:
\[
  |V| \cdot (|V| - 1) \leq C_{100\%} < |V| \cdot (|V|^2 - 1).
\]

\proof{} The lower bound is obtained similarly as in
\lemmaref{elemsrepcomplete}. The least amount of communication to achieve full synchronization is equivalent to each node sending its element to all the other nodes directly. On the other hand, we get the upper bound by observing that there cannot be more than
$|V|^{2} \cdot (|V| - 1)$ \emph{redundant} element transmissions on top of the lower bound. Redundant transmissions happen when a node receives an element via multiple replicas in the same iteration. 
To count all redundant transmissions, we observe that, in each iteration, each node either receives some new elements or does not receive any. In the latter case, obviously, no redundant transmissions happen. Otherwise, if there are some new elements received, the
following holds:
\begin{enumerate*}[label=(\arabic*)]
\item there will be no more than $|V|$ new elements arriving at the
  node across all iterations, as there is only that much elements in the
  network, and
\item for each element, there cannot be more than $|V| - 1$ redundant
  transmissions, as there cannot be more than that much replicas at
  any node.
\end{enumerate*}
Thus, there cannot be more than $|V|^{2} \cdot (|V| - 1)$ redundant
transmissions at all nodes in all iterations.\qqed{}

As in Watts-Strogatz networks we have $\overline{deg}$ replicas at
each node on average, the same counting argument from above applies in
the following form.

\coroll{comm_ws} For \epsrep{} in Watts-Strogatz networks:
\[
  C_{100\%} < |V| \cdot (|V| \cdot \overline{deg} + |V| -
  1).
\]

On the other hand, to infer the upper bound on the time that \epsrep{}
needs to complete a full sync ($T_{100\%}$), we rely on following
definition.

\definition{iteration} $I_{x\%}(G)$ is the maximal number of \epsrep{}
iterations (\lineref{line:srep_main_loop} in
\aref{algo:parallel_srep}) at any node to achieve x\% network
synchronization.

\theorem{iter_bound} In \epsrep{} and for any connected network
$G = (V, E)$, with the shortest path between nodes $u$ and $v$ denoted
as $dist(u, v)$, the maximum number of iterations required for a full
network synchronization is equal to the diameter of the network:
\[
  I_{100\%}(G) = \max_{u, v \in V} dist(u, v).
\]

\proof{} By the definition of full synchronization, all elements need
to reach every other node. Without a loss of generality, suppose that
we follow the propagation of some element $i \in V$ during the
execution of \epsrep{}. Since the graph is connected, in each
iteration of \epsrep{}, $i$ will progress exactly one step further
through the network. The number of iterations required to synchronize
the entire network is then equivalent to the maximum distance between
any two nodes in the network (\ie diameter).\qqed{}

\lemma{par_srep_compl} In \epsrep{} over complete graphs $G = (E, V)$:
\[
  I_{100\%}(G) = 1 \text{ and } C_{100\%} = |V| \cdot (|V| - 1).
\]
The former holds as the diameter of complete graphs is 1. The latter
is a consequence of the former; as no element traverses more than one
edge, there cannot be any redundant transmissions.

\coroll{iter_ws} For \epsrep{} and Watts-Strogatz networks, the
maximal number of iterations at any node to synchronize the entire
network ($I_{100\%}$) is logarithmic in the size of the network.

Counting the number of nodes that have heard about an element
$n \in V$ in iteration $i$ of \epsrep{} over a Watts-Strogatz network,
we get the following sum:
\[
  1 + \overline{deg} + \overline{deg}^2 + \ldots + \overline{deg}^{i}.
\]
By equating it to $|V|$, we can express $i$, the number of iterations
until all nodes have heard of $n$, as a logarithmic function of
$|V|$~\cite{chung2001diameter}. Practically speaking, \epsrep{} will
complete in logarithmically small number of iterations
($\approx 4\log_{\overline{deg}}(10))$) for the blockchain networks of
realistic sizes (\eg{} Blockchain and
Ethereum~\cite{txprobe,degwithunreachable}).

\theorem{par_srep_time} In general graphs $G = (V, E)$, the following
holds for \epsrep{}:
\begin{align*}
  & T_{100\%} \leq I_{100\%}(G) \cdot \max_{i \in V} t_{i} < I_{100\%}(G) \cdot
    |V|,\\
  & \Sigma_{100\%} \leq I_{100\%}(G) \cdot |E|.
\end{align*}

\proof{} Since synchronizations happen in parallel, the overall
elapsed time is proportional to the number of iterations. Any sync
invocation at any node will take strictly less than $|V|$, as no two
data sets can differ in more than $|V| - 1$ elements (each data set
keeps exactly one element at the beginning). Since in each iteration
nodes sync with all their neighbors and each sync is two-way by
\ref{a:two_way}, there will be no more than $|E|$ syncs in each
iteration.\qqed{}


\begin{figure}
  \centering
  \includegraphics[width=.9\columnwidth]{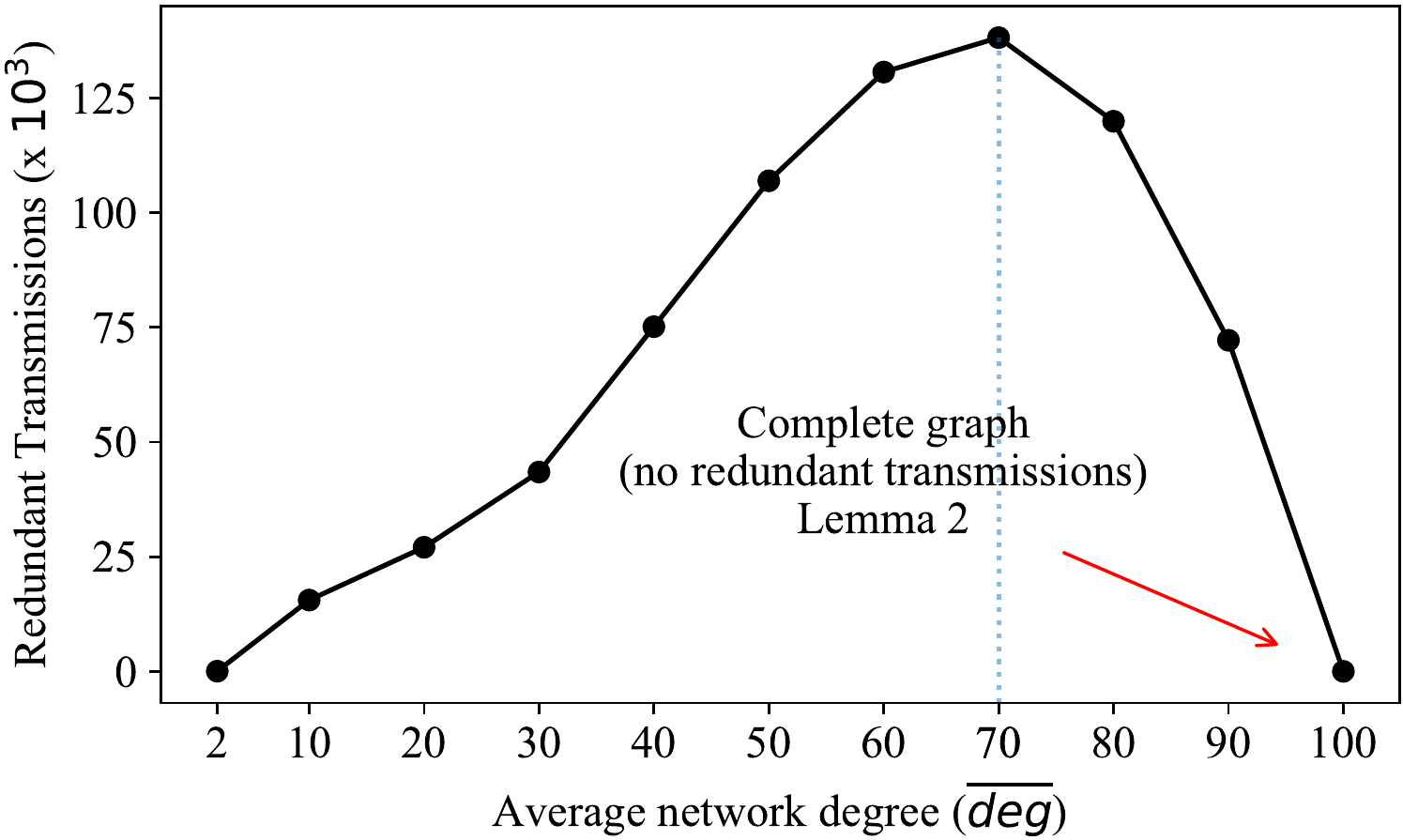}
  \caption{Amount of redundant transmissions in \epsrep{} over a
    network of 100 nodes ($p = 0.24$).} \label{fig:redundant}
\end{figure}

\subsubsection*{The $\overline{deg}$ Dilemma}
Due to the counting argument from \theoremref{comm_bounds}, the upper
bound on overall communication cost is \emph{not} tight; there must be
at least some elements that will \emph{not} generate redundant
transmissions in any connected network. 
On top of that, the topology of the network plays a complex role in
generating redundant transmissions. Intuitively speaking, the impact
of $\overline{deg}$ in Watts-Strogatz networks is twofold, and
conflicting:
\begin{enumerate*}[label=(\arabic*)]
\item the larger $\overline{deg}$, the larger the average number of
  replicas per node, which may cause redundant transmissions, and
\item the larger $\overline{deg}$, the shorter the average pair-wise
  shortest path among the nodes in the network, which makes each
  element traverse less intermediate nodes to reach the entire network,
  thus reducing the probability
  of redundant transmissions.
\end{enumerate*}
We plot this non-monotonic effect that $\overline{deg}$ has on the
amount of redundant transmissions in \fref{fig:redundant} for a
tractably small network. Up to a point, the first effect (replicas
count) prevails and drives the overall communication cost up. After
that point, the second effect (path shortening) prevails and drives
the overall communication cost down all the way to the point when the
network becomes a complete graph and there is no redundant
transmissions at all.

\subsection{Multi-element \srep{}}\label{sec:me_srep}
%

The final stage in building \srep{} is \emph{multi-element
  \srep{}}. We build it by relaxing \ref{a:single_elem} ---
transaction pools can now initially contain multiple elements. In
terms of our network model, this means that our $\mathcal{S}$ (sizes
distribution) and $\mathcal{P}$ (differences distribution) are no more
constant.
Thus, \srep{} is a \emph{generalization} of \epsrep{}.

\definition{f_func} Function $f: (G, A) \mapsto \mathbb{Z} $ maps a
pair of a topology $G$ and a pool assignment $A$ to a non-negative
integer via first constructing the corresponding mutual differences
matrix $M$, then computing $\sum m_{ij}$.

\definition{g_func} Function $g: (G, A) \mapsto (G, A_{(next)})$ maps a pair
of a topology $G$ and a pool assignment $A$ to the same topology $G$
and a transformed pool assignment $A_{(next)}$. We define the transaction
pools in the transformed pools assignment $A_{(next)}$ as:
\[
  S_{(next)i} = S_{i} \cup (\bigcup_{j \in G[i]} S_{j}).
\]

We use $\bigcup_{j \in G[i]} S_{j}$ to denote the union of all
transaction pools $S_j$ corresponding to the neighbors of node $i$ in
the previous iteration.

\definition{comp_with_itself} For some function $h$, we write
$h^{(n)}(x)$ to denote the composition of function $h$ with itself
$n$ times, starting with argument $x$:
\[
  h^{(n)}(x) = \underbrace{h \circ h \cdots h}_{n} (x).
\]

\definition{asgn_g_foo} $A_{(n)}$ is the assignment resulting from $n$
compositions of $g$ with itself starting with the initial pool
assignment that we denote as $A = A_{(0)}$.

\lemma{me_srep} For a network model $(G, A)$ where $G$ is a
connected graph and $A$ the initial pool assignment,
the number of \srep{} iterations to achieve the full network
synchronization $I_{100\%}(G, A)$ is given as a solution to the
following equation:
\[
  f(g^{(I_{100\%}(G, A))}(G, A)) = 0.
\]


Note that by \defref{g_func}, $g$ exactly corresponds to one iteration
of \srep{}. That is, the transformed pool assignment $A_{(next)}$
reflects the state of the transaction pools after an iteration of
\srep{} at all nodes in the network. Composing $g$ with itself $n$
times corresponds to repeating an iteration of \srep{} at all nodes
$n$ times. By a similar argument as in \theoremref{iter_bound}, all
elements will reach all nodes after some number of iterations. Since
this implies that no two sets have any differences, $M$ will be an
all-zeros matrix. That is, $(f \circ g^{(n)}) (G, A)$ has at least one
zero. Thus, the number of times we need to compose $g$ with itself
until $f(G, A_{(n)}) = 0$ gives us the maximal number of \srep{}
iterations to achieve full network synchronization.

\theorem{me_srep_iter} For a connected graph $G = (V, E)$ and an
initial pool assignment $A$, the number of \srep{} iterations to
achieve the full network synchronization is bounded by the diameter of
the network:
\[
  I_{100\%}(G, A) \leq \max_{u, v \in V} dist(u, v).
\]

\proof{} As \srep{} is a generalization of \epsrep{}, the argument
here is similar to that of \theoremref{iter_bound}. To achieve the
full network synchronization, elements need to traverse at most the
diameter of $G$. As opposed to \epsrep{}, in \srep{} each element may
initially appear at more than one node, dictated by the differences
distribution $\mathcal{P}$. Thus the diameter is an upper bound on
\srep{} iterations.\qqed{}

\lemma{me_srep_comm} For a connected graph $G = (V, E)$ and initial
pool assignment $A$ with the corresponding mutual differences matrix
$M$, the communication cost of \srep{} is:
\begin{align*}
  &C_{100}(G, A) = \sum_{i = 0}^{I_{100\%}(G, A)} f(G, A_{(i)}) \\
                & < I_{100\%}(G, A) \cdot \max\{f(G, A),\dotsc,f(G,
                  A_{(I_{100\%(G, A)})})\}.
\end{align*}

In $i$th iteration of \srep{}, we transmit exactly as much elements
as there are in the differences matrix that corresponds to
$A_{(i)}$. Given $I_{100\%}(G, A)$ from \lemmaref{me_srep}, we get the
overall communication cost of \srep{}.


\lemma{me_srep_inv} In \srep{} over a connected network
$G = (V, E)$ with the given initial pool assignment $A$ and the
largest order statistics of differences distribution $\mathcal{P}$
denoted as $\mathcal{P}_{(n)}$:
\begin{align*}
  & T_{100\%} \leq I_{100\%}(G, A) \cdot \max_{i \in V} t_{i} =
    I_{100\%}(G, A) \cdot \mathcal{P}_{(n)}, \\
  & \Sigma_{100\%} \leq I_{100\%}(G, A) \cdot |E|.
\end{align*}

The argument is similar to that of \theoremref{par_srep_time}.












Finally, note that the assumptions in our analysis such
as~\ref{a:data_gen} --- no new transactions arrive after \srep{}
starts, are artificial in that they simplify our analysis, but they do
not constrain \srep{} in practice. The properties such as the overall
communication cost ($C_{100\%}$) and time ($T_{100\%}$) to sync the
entire network relate to the transactions that have arrived before
\srep{} begins.

\section{Simulations}\label{sec:sim}
To validate our analytical findings about \srep{}, we construct an
event-based simulator called \srepsim{}~\cite{srepsim_code} that
shares the topology generation procedure with \emph{CBlockSim} of Ma
\ea{}~\cite{cblocksim} and adds the other parameters of our network
model described in \sref{sec:net_model}.


In the rest of this section, we first describe a method to parameterize our network model. Then, we use such parameterized model to validate the main analytical properties of \srep{}. We then compare
the overall communication cost of \srep{} with a similar approach from
the literature. At the end, we present a \srepsim{} optimization that
allows for easy \srep{} communication cost calculation over
large-scale networks.

\subsection{Configuring Network Model
  Parameters}\label{sec:config_net}
Unlike the simulation approaches from the literature (\eg{}
\emph{SimBlock}~\cite{simblock}), our network model can seamlessly
integrate real-world transaction pool data. For instance, the
empirical
distributions of $\mathcal{S}$ and $\mathcal{P}$ can be generated for
some small subset of all nodes in the network using the measurement
software such as \emph{log-to-file} of Imtiaz
\ea{}~\cite{anas_orphan_icbc,anas_orphan_tnsm}. This software
instruments adjacent Bitcoin nodes and periodically serializes the
snapshots of their transaction pools. From these transaction pool
snapshots, we can measure transaction pool sizes and their mutual
differences to construct the empirical distributions for $\mathcal{S}$
and $\mathcal{P}$.

\begin{figure}
  \centering
  \includegraphics[width=.8\columnwidth]{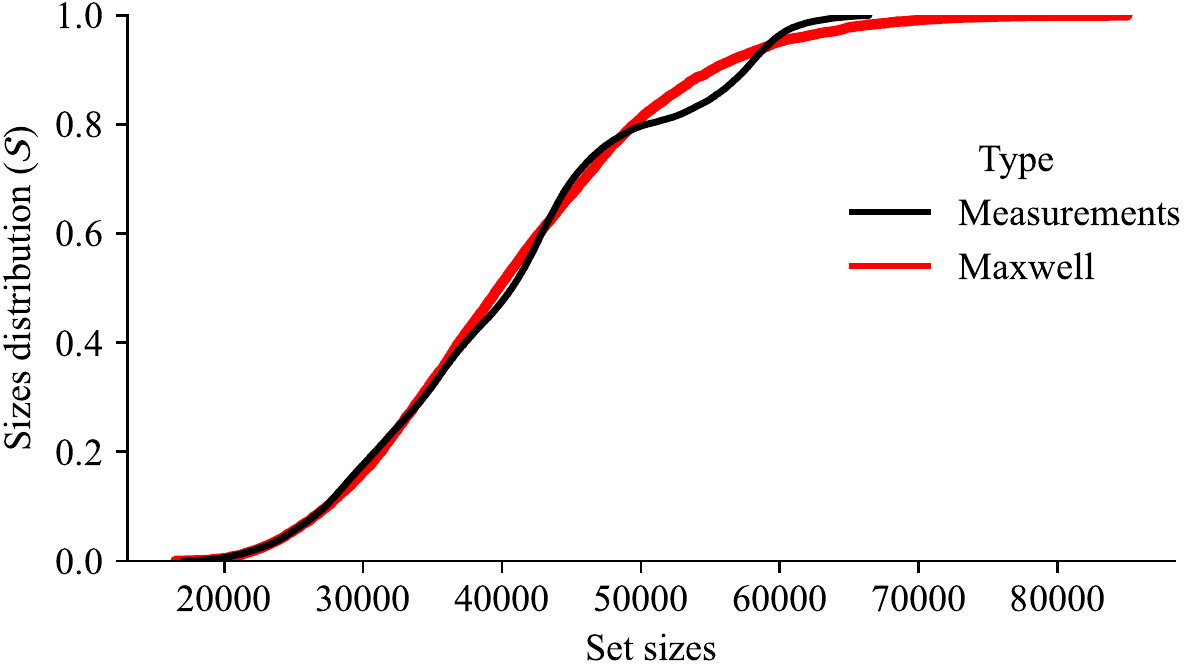}
  \includegraphics[width=.8\columnwidth]{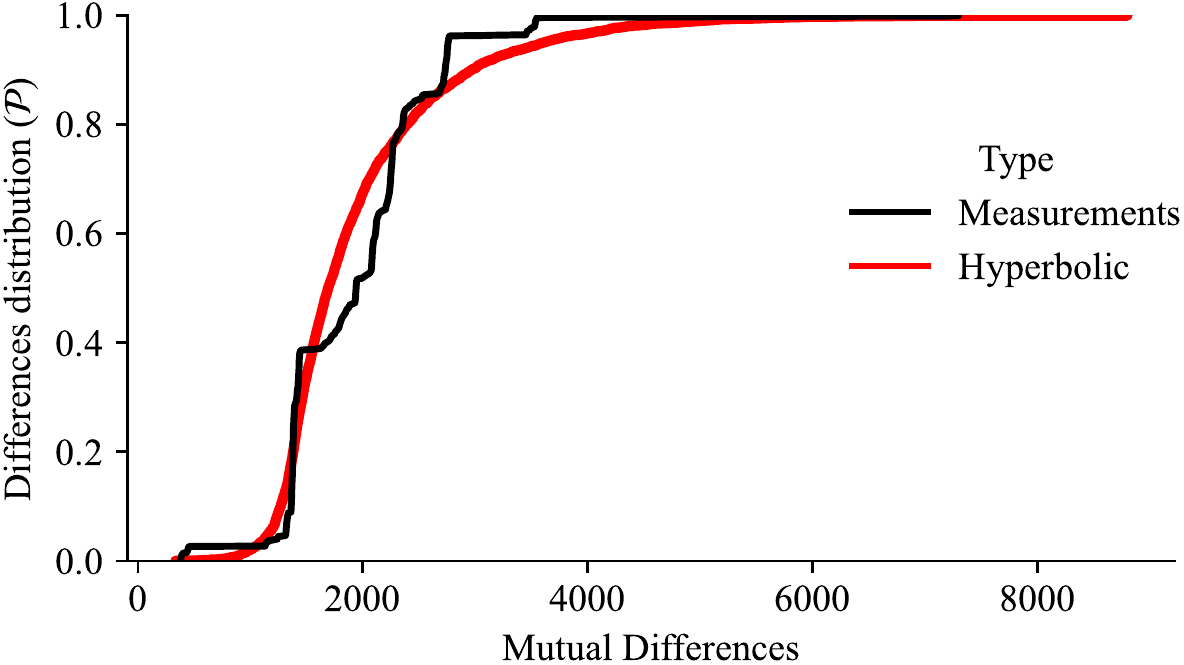}
  \caption{Empirical distributions of transaction pool sizes
    $\mathcal{S}$ for two adjacent Bitcoin nodes (up) and their mutual
    differences $\mathcal{P}$ (down). Best distribution fits in red
    (using Error Sum of Squares).}
  \label{fig:diffs_dist}
\end{figure}

For the purpose of this work, we have conducted a 3-day long
measurement campaign on two time-synchronized Bitcoin nodes and
requested the transaction pool snapshots each
minute. \fref{fig:diffs_dist} depicts the results that we
obtained. Roughly speaking, the set sizes fit the Maxwell distribution
reasonably well, while the number of mutual differences fits the
Hyperbolic distribution. Next, given the empirical distribution of
$\mathcal{S}$, we need to configure the rest of our network model's
pool parameters\footnote{Direct usage of $\mathcal{P}$ is also
  possible but perhaps harder. 
  }. Ultimately, we need to construct a pool
  assignment $A$ that conforms to the differences distribution
$\mathcal{P}$.

\begin{figure}
  \centering
  \includegraphics[width=.9\columnwidth]{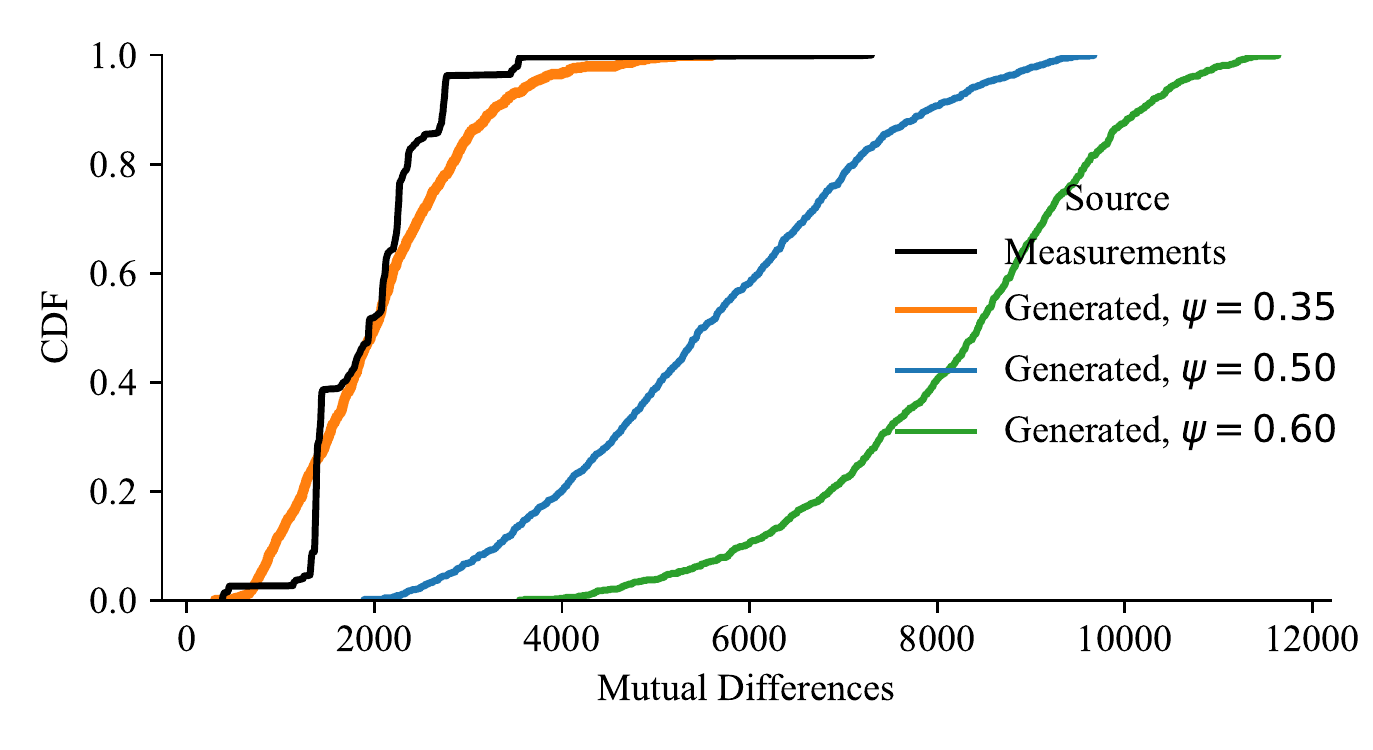}
  \caption{Empirical differences distribution for two adjacent Bitcoin
    nodes versus the differences distribution generated by
    \procref{proc:a_gen} for various $\psi$. Watts-Strogatz network
    with 100 nodes ($\overline{deg} = 19$ and $p = 0.24$).}
  \label{fig:psi}
\end{figure}

In \srepsim{}, we construct such assignments through
\procref{proc:a_gen}. For the given network topology $G = (V, E)$ and
the sizes distribution $\mathcal{S}$, we need to configure the
parameter $\psi$ such that the resulting assignment $A$ produces a
differences distribution that resembles $\mathcal{P}$. As shown in
\fref{fig:psi}, $\psi = 0.35$ works reasonably well with our empirical
sizes distribution. Note that by increasing $\psi$, we can decrease
the average similarity among the transaction pools (\ie{} increase the
number of their mutual
differences).

\begin{procedure}
  \small
  \caption{Network parameterization in SREPSim.()}\label{proc:a_gen}
  \KwIn{Network $G = (V, E)$.}
  \KwIn{Sizes distribution $\mathcal{S}$.}
  \KwIn{Parameter $\psi$.}
  \KwOut{Pool assignment $A$.}
  \SetKw{kwSample}{sample}
  \SetKw{kwElemsFrom}{elements from}
  \SetKw{kwForIn}{in}
  $u \gets \lceil\ {\psi \E[\mathcal{S}]}\ \rceil$ \;
  $\mathcal{U}\{0, u - 1\}$ \tcp*{Uniform distribution}
  sizes $\gets$ \kwSample $|V|$ \kwElemsFrom $\mathcal{S}$ \;
  A $\gets$ [ ] \;
  \For{$i \gets 0$ \KwTo $|V| - 1$} {
    $S_{i} \gets$ \kwSample sizes[i] \kwElemsFrom $\mathcal{U}$ \;\label{proc:a_gen:line_col}
    A.append ( $S_{i}$ ) \;
  }
\end{procedure}

\subsection{\srep{} Properties Validation}
The main analytical properties that we want to validate through
simulations are \srep{}'s communication cost to achieve full network
sync ($C_{100\%}$) and the time required to achieve this state
($T_{100\%}$). In particular, we want to show how these two quantities
change as a function of the network topology and the measure of
difference among the transaction pools. 

\begin{figure}
  \centering
  \includegraphics[width=.8\columnwidth]{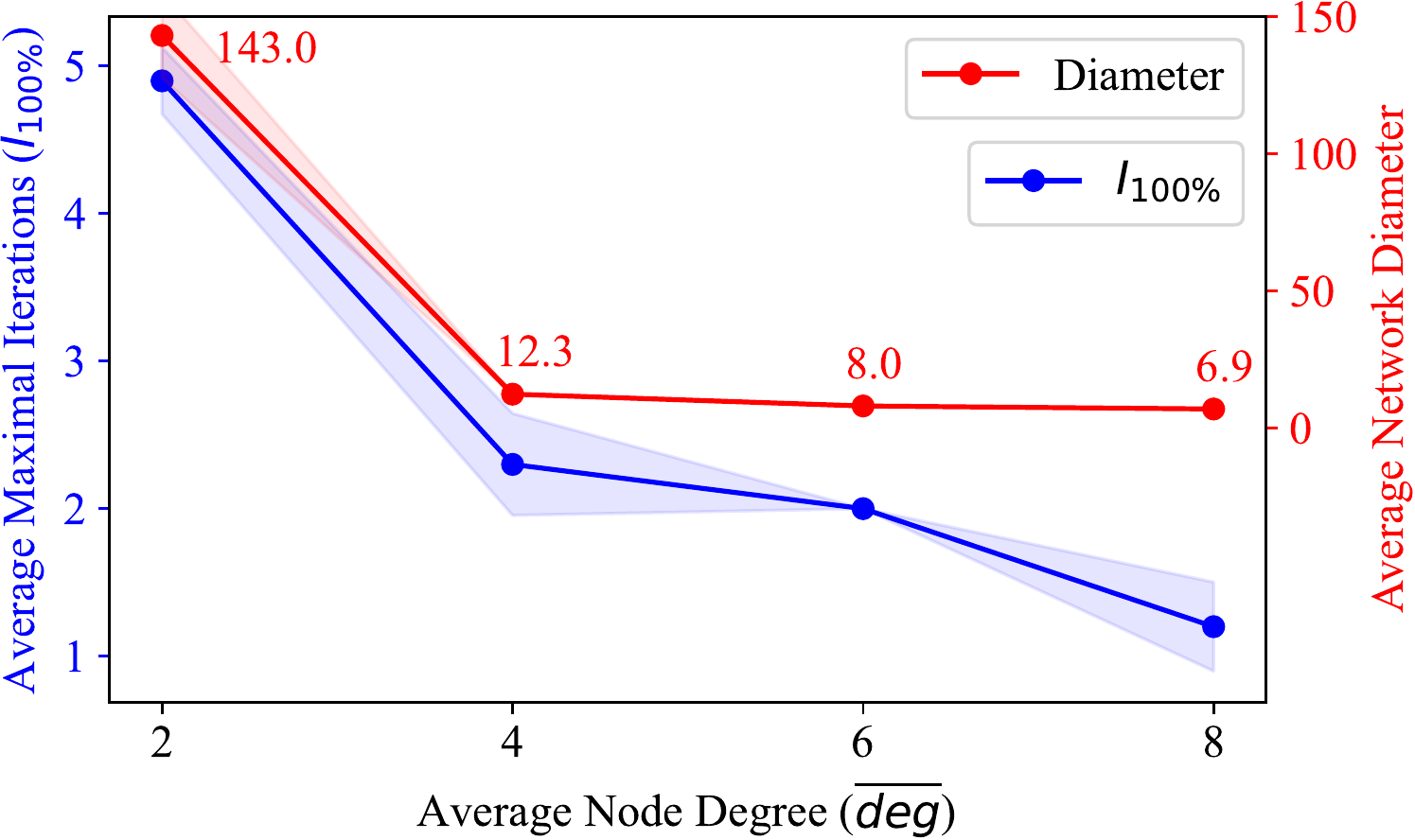}
  \caption{Maximal number of \srep{} iterations at any node
    ($I_{100\%}$) bounded by the network diameter for Watts-Strogatz
    graphs with 1000 nodes ($p = 0.24$). 95\% confidence intervals.}
  \label{fig:i_vs_diam}
\end{figure}

\begin{figure}
  \centering
  \includegraphics[width=.8\columnwidth]{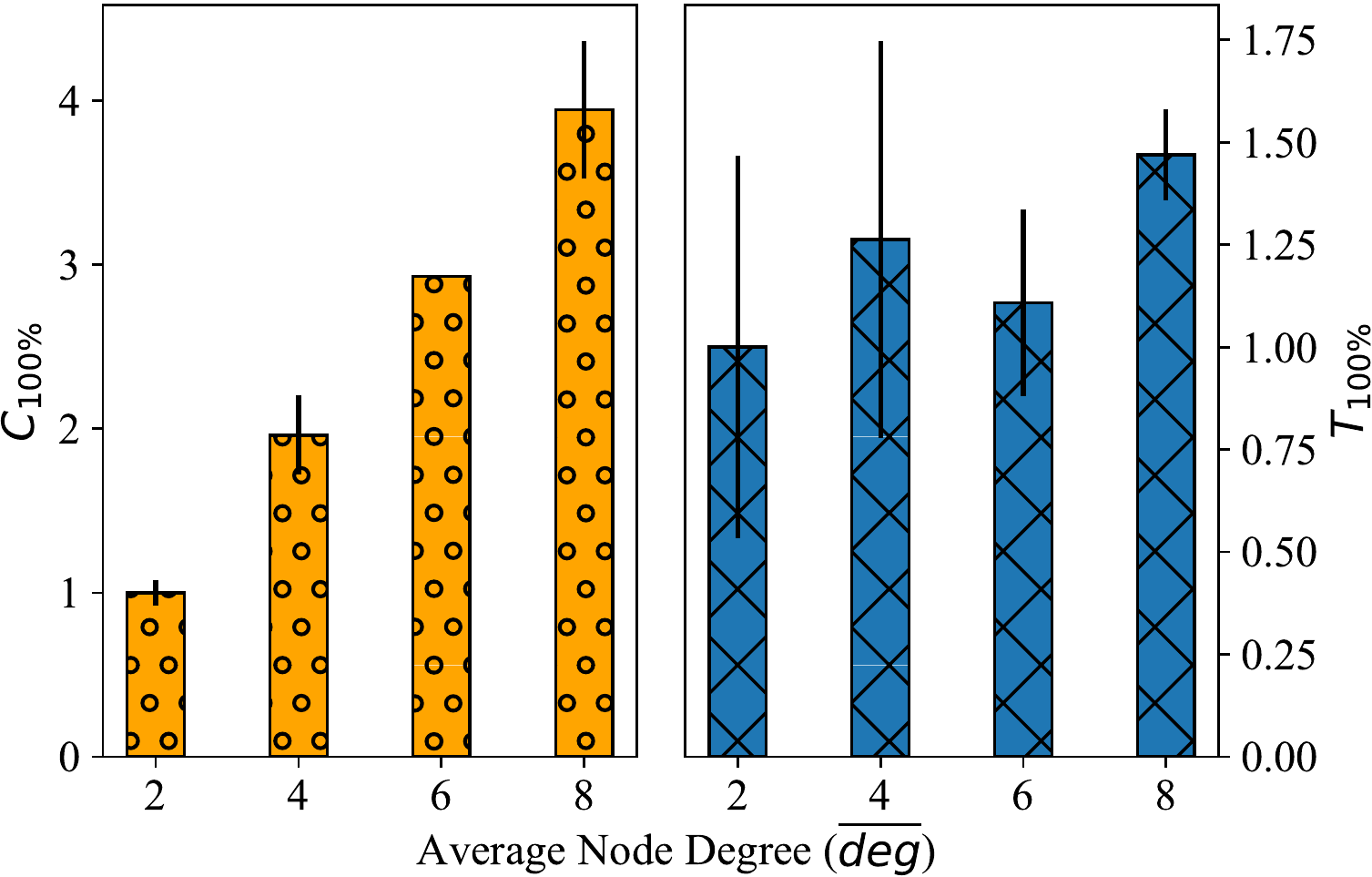}
  \caption{Relative communication cost ($C_{100\%}$) and time to fully
    synchronize the network ($T_{100\%}$). Network with 1000 nodes
    ($p = 0.24$).}
  \label{fig:comm_and_t}
\end{figure}

In \fref{fig:i_vs_diam}, we plot the maximal number of \srep{}
iterations $I_{100\%}$ and the network diameter as functions of the
average network degree $\overline{deg}$. In \fref{fig:comm_and_t}, we
plot the communication cost and time to full network sync as a
function of $\overline{deg}$. The main observation is that the overall
communication increases with the average node degree as a consequence
of using more replicas per node, which increases the number of
redundant transmissions (see \fref{fig:redundant}). On the other hand,
the time to achieve full network sync does not exhibit such a
trend. Since primal syncs run in parallel, it is the maximal number of
differences among any two nodes in the network that dominates the
total time to sync the network (see \lemmaref{me_srep_inv}).

\subsection{Comparison with \mempoolsync{}}\label{sec:mempoolsync}

\mempoolsync{} of Imtiaz \ea{} is a transaction pool synchronization
protocol that can improve the average transaction propagation delay by
50\% in the event of churn in the Bitcoin
network~\cite{anas_churn_tnsm}. Here we describe this protocol and
compare its communication efficiency with our newly proposed \srep{}
through simulations.

As pointed out in~\cite{anas_churn_tnsm}, the main reason
for slow block propagation times is a large number of missing
transactions in the transaction pools of the block-receiving
nodes. This effect occurs in the legacy block propagation protocols
such as \emph{CompactBlock}~\cite{compact_block} and the more recent
improvements such as
\emph{Graphene}~\cite{ozisik2019graphene,anas_empir}. Thus, the goal
of \mempoolsync{} is to supply the nodes with potentially missing
transactions, and it does so through an \emph{ancestor score}-based
heuristics~\cite{bitcoin_ancestor_score}. The protocol uses a small
constant \texttt{DefTXtoSync} as the default number of transaction
hashes that the transmitting node will select from its transaction
pool in descending order of ancestor score. The transmitting node will
send exactly \texttt{DefTXtoSync} selected transaction hashes
\emph{unless} one of the following holds:
\begin{enumerate}[label=\arabic*)]
\item Transmitting node's transaction pool is much larger than
  \texttt{DefTXtoSync} (\eg{} 10 times). In this case, the node
  will send $Y \times \text{\texttt{DefTXtoSync}}$ top rated
  transactions, where $Y$ is a constant between 0 and 1, or
\item Transmitting node's transaction pool is smaller than
  \texttt{DefTXtoSync}. In this case, the node will send its entire
  transaction pool. Because \texttt{DefTXtoSync} is a small constant,
  this is a quite rare event. It occurs only when the node has just
  joined the Bitcoin network or has just propagated a large block that
  triggered a massive transaction pool cleanup~\cite{anas_churn_tnsm}.
\end{enumerate}

\begin{figure}
  \centering
  \includegraphics[width=.8\columnwidth]{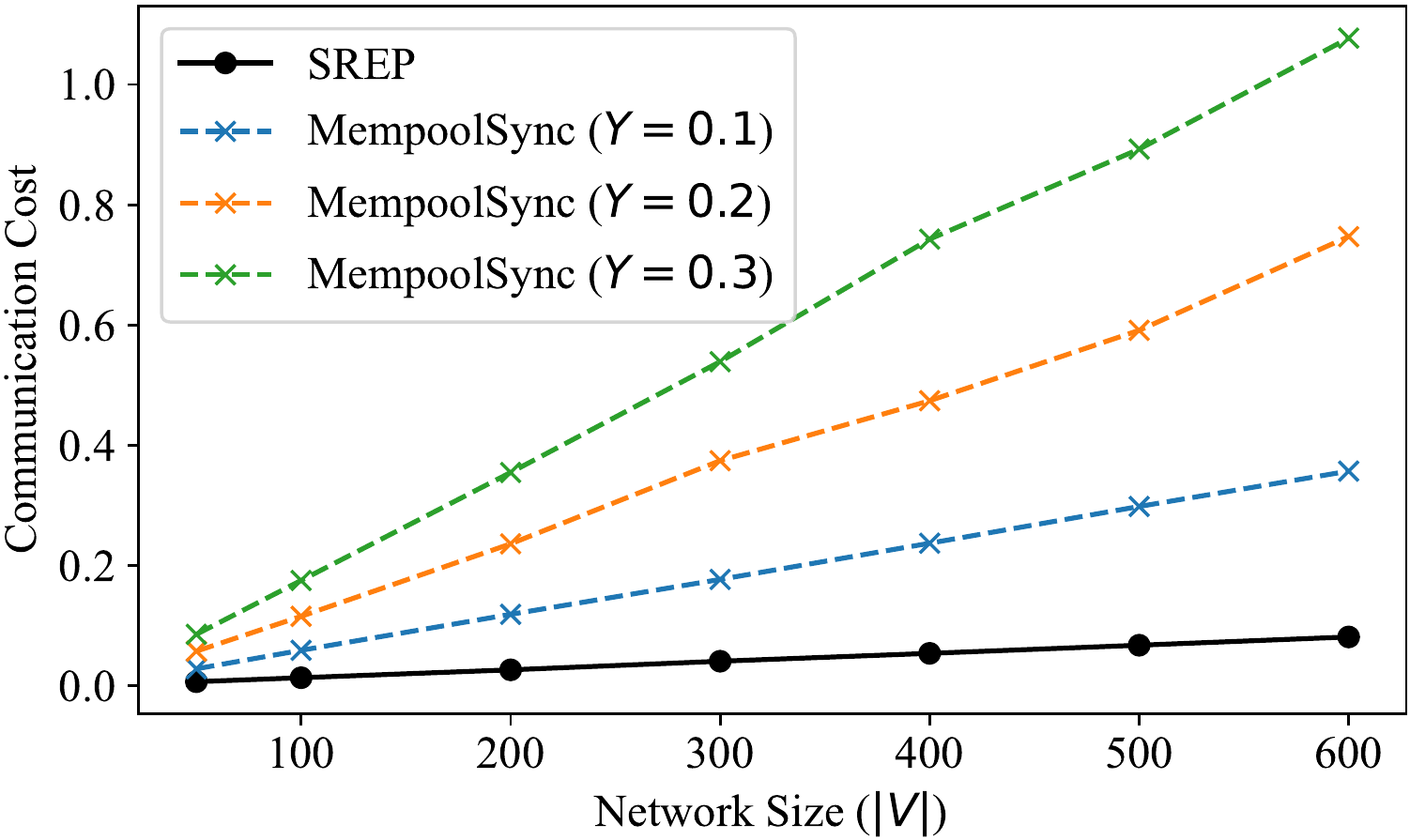}
  \caption{Normalized overall communication cost of \srep{}
    ($C_{100\%}$) and \mempoolsync{} as a function of network
    size. Data from~\sref{sec:config_net}. $DefTXtoSync = 1000$. $Y$
    is the \mempoolsync{} heuristic constant.}
  \label{fig:mempool}
\end{figure}

In \fref{fig:mempool}, we compare the overall communication costs of
\mempoolsync{} and \srep{}. For \srep{}, we plot the communication
cost to sync the entire network $(C_{100\%})$. For \mempoolsync{}, we
plot the communication cost that \mempoolsync{} incurs until \srep{}
would achieve a full sync.

Note that this kind of comparison gives an advantage to
\mempoolsync{}. While \srep{}'s $C_{100\%}$ implies that the network
is fully synced, \mempoolsync{}'s communication cost does not. In
fact, \mempoolsync{} has no guarantees about the communication (or
time) needed to sync the entire network. Note also that \mempoolsync{}
uses Bitcoin internals to calculate the ancestor score of the
transactions and later uses this score to determine which transactions
to transmit. As opposed to \mempoolsync{}, \srep{} is a general
approach that does not rely on any Bitcoin internals and can be
seamlessly integrated into other blockchains that keep transaction
pools.


\begin{table}
  \centering
  \begin{tabular}{>{\centering\arraybackslash}p{.5cm}>{\centering\arraybackslash}p{1cm}>{\centering\bfseries\arraybackslash}p{1cm}>{\centering\bfseries\arraybackslash}p{1cm}>{\centering\bfseries\arraybackslash}p{2cm}}
    \toprule
    $\overline{deg}$ & $\psi$ & Diameter \textit{average} & $I_{100\%}$ \textit{average} & $C_{100\%}$ (GB) \textit{average} \\
    \midrule
    \multirow{3}{*}{4} & 0.355 & \multirow{3}{*}{16} & 2.5 & 1.214397\\
                     & 0.5 & & 3.0 & 3.165879\\
                     & 0.6 & & 3.1 & 4.801665\\
    \midrule
    \multirow{3}{*}{8} & 0.355 & \multirow{3}{*}{9} & 1.7 & 2.428649\\
                     & 0.5 & & 2.0 & 6.317304\\
                     & 0.6 & & 2.0 & 9.569259\\
    \midrule
    \multirow{3}{*}{12} & 0.355 & \multirow{3}{*}{7} & 1.0 & 3.642738\\
                     & 0.5 & & 1.5 & 9.485572\\
                     & 0.6 & & 2.0 & 14.347242\\
    \midrule
    \multirow{3}{*}{16} & 0.355 & \multirow{3}{*}{6} & 1.0 & 4.876714\\
                     & 0.5 & & 1.0 & 12.649385\\
                     & 0.6 & & 1.0 & 19.135943\\
    \midrule
    \multirow{3}{*}{20} & 0.355 & \multirow{3}{*}{5} & 1.0 & 6.065679\\
                     & 0.5 & & 1.0 & 15.804836\\
                     & 0.6 & & 1.0 & 23.886079\\
    \midrule
    \multirow{3}{*}{24} & 0.355 & \multirow{3}{*}{5} & 1.0 & 7.294909\\
                     & 0.5 & & 1.0 & 18.966694\\
                     & 0.6 & & 1.0 & 28.672272\\
    \midrule
    \multirow{3}{*}{28} & 0.355 & \multirow{3}{*}{5} & 1.0 & 8.465624\\
                     & 0.5 & & 1.0 & 22.156316\\
                     & 0.6 & & 1.0 & 33.446278\\
    \bottomrule
  \end{tabular}
  \caption{\srep{} over a 10,000 nodes network. $p = 0.24$.}
  \label{tab:srep_large}
\end{table}

\subsection{Communication Cost in Large-Scale
  Networks}\label{sec:large_net}

Event-based simulators such as \srepsim{} may consume prohibitive
amounts of memory and take a long time to complete simulations when
the simulated network is large~\cite{cblocksim}. To address this
issue, we designed a \srepsim{} module that computes \srep{}'s
performance metrics analytically. In particular, we implement the
functions from Definitions~\ref{definition:f_func}
and~\ref{definition:g_func}, and rely on the results from
\lemmaref{me_srep_comm} to compute $C_{100\%}$ and $I_{100\%}$. We
describe the \srepsim{}'s analytical module in
\procref{proc:comm_analytic}. Using this module, we can easily compute
the desired performance metrics for the networks of realistic sizes
(\eg{} Bitcoin and
Ethereum)~\cite{txprobe,degwithunreachable}.

In~\tref{tab:srep_large}, we summarize the results for a 10,000 nodes
network with various average node degrees ($\overline{deg}$) and the
measure of similarity among transaction pools ($\psi$). As we report
the communication cost, we assume that the transaction pools represent
each transaction as a 32-byte long globally unique
hash~\cite{utxo}. All simulations complete in tens of minutes.

\begin{procedure}
  \small
  \caption{SREPSim's analytical module.()}\label{proc:comm_analytic}
  \KwIn{Network $G = (V, E)$.}
  \KwIn{Initial pool assignment $A$ as $S_0..S_{|V| - 1}$.}
  \KwOut{Overall network communication cost $C_{100\%}$.}
  \KwOut{Maximal number of iterations $I_{100\%}$.}
  \SetKwFunction{FCalcM}{CalculateM}
  \SetKwProg{Fn}{function}{:}{}

  \Fn{\FCalcM{$A$}}{
    $M \gets$ zeros($|V| \times |V|$) \tcp*{Zero matrix}
    \For{$i \gets 0$ \KwTo $|V| - 1$} {
      \For{$j \gets i + 1$ \KwTo $|V| - 1$} {
        \uIf(\tcp*[h]{i neighbor of j}){$i \in G[j]$} {
          $M[i][j] \gets |S_i \oplus S_j|$ \;
        }
      }
    }
    \KwRet $M$\;
  }

  $C_{100\%} \gets 0$ \;
  $I_{100\%} \gets 0$ \;
  $M \gets$ \FCalcM($A$) \;
  \While{$\sum m_{ij} > 0$} {
    \For{$i \gets 0$ \KwTo $|V| - 1$} {
      $S'_{i} \gets S_{i}$ \tcp*{New assignment}
      \For{$j \in G[i]$} {
        $S'_{i} \gets S'_{i} \cup S_{j}$ \;
      }
    }
    $C_{100\%} = C_{100\%} +  \sum m_{ij}$ \;
    $I_{100\%} \gets I_{100\%} + 1$ \;
    $A \gets A'$ \;
    $M \gets$ \FCalcM($A$) \;
  }
\end{procedure}

\section{Conclusion}\label{sec:conclusion}
In this work, we have developed and analyzed \srep{}, an independent
protocol that assists block propagation in large-scale blockchains.
This new protocol synchronizes transaction pools of nodes in the
blockchain network using communication-efficient set reconciliation
approaches from the literature.  However, rather than inserting itself
directly into the block propagation process, as previous works have
done, \srep{} operates in a distributed manner \emph{outside} the
block propagation channels of the network.  As a result, it is easier
to formally analyze its performance, and, indeed, we have shown that
it completes in time bounded by the network diameter (or logarithmic
in network size for the ``small-world'' networks that reasonably model
blockchain networks).

We have also validated our analytical findings against a novel
event-based simulator that we have developed.  We run the simulator on
real-world transaction pool statistics drawn from our own measurement
campaign.  In our simulations, \srep{} incurs only tens of gigabytes
of overall bandwidth overhead to synchronize networks with ten
thousand nodes, which is several times better than the current
approach in the literature.


For future work, we propose to consider \emph{multi-party} set
reconciliation~\cite{multiparty,multiparty_cpi} in the context of
transaction pool sync. Though the main benefit may be further
reduction in overall communication cost, it is not clear whether an
advantage over pairwise approaches can be achieved when an average
pairwise intersection is large compared to the total intersection
($\cap_{i}S_{i}$)~\cite{multiparty}.


\vspace{-5pt}
\section*{Acknowledgments}
The authors would like to thank Red Hat, the Boston University Red Hat
Collaboratory (award \# 2022-01-RH03), and the US National Science
Foundation (award \# CNS-2210029) for their support.

\bibliographystyle{IEEEtran}
\bibliography{bibliography}


\end{document}